\documentstyle[epsfig,aps]{revtex}

\begin{document}

\twocolumn[\hsize\textwidth\columnwidth\hsize\csname@twocolumnfalse\endcsname

\title{Infinite series of magnetization plateaus in the frustrated
dimer-plaquette chain
}
\author{J. Schulenburg and J. Richter }
\address{
Institut f\"ur Theoretische Physik, Universit\"at Magdeburg, P.O.Box 4120, 
D-39016 Magdeburg, Germany
}

\date{\today}
\maketitle

\begin{abstract}
The dimer-plaquette chain undergoes a first order transition driven by
frustration to an orthogonal  dimer ground state. We analyze the
magnetization curve in this orthogonal dimer regime. Besides of the recently
found magnetization plateaus at $m=1/4$ and the $m=1/2$  we find an infinite 
sequence of
plateaus between $m=1/4$ and $m=1/2$. These additional 
plateaus
belong to magnetizations $m(n)=n/(2n+2)$, where $n$ is a positive 
integer number.
\end{abstract}

]  


The frustrated dimer-plaquette chain (or orthogonal dimer chain)
introduced in \cite{ivanov} has
attracted attention because of its first order quantum phase 
transition to a product dimer ground state and its close relation to the
two-dimensional spin model for SrCu$_2$(BO$_3$)$_2$ \cite{ueda}.      
Motivated by the recent discovery of magnetization plateaus in SrCu$_2$(BO$_3$)$_2$ 
Koga {\it et.\ al.} \cite{koga00} investigated 
the magnetization curve for the one-dimensional model.
They found 
magnetization plateaus at $m=1/4$ and $m=1/2$ of the full moment 
near the phase transition point 
using exact diagonalization and DMRG methods.
In order to reproduce their results we studied the magnetization curve using
large-size exact diagonalization (up to N=40 sites) and analytical
considerations based on the product nature of a certain class of
eigenstates of the orthogonal dimer chain \cite{ivanov}. We found clear 
evidence of 
further magnetization
plateaus between $m=1/4$ and $m=1/2$. 
In what follows we will explain our results for  parameters $J=1$ and 
$j=J'/J=0.7$ 
(for notations see Fig.~1 and Refs. \cite{ivanov,koga00}), 
i.e., in the zero-field regime the ground state is a
dimer product state,  
but the argumentation is more general and can be applied in principle 
to the (b)-phase of the phase diagram presented in Fig.~9 of \cite{koga00}.

To show that we have not only a single jump between 
the $m=1/4$ and $m=1/2$ plateaus we need two facts.

First we have to know the lowest energy of eigenstates with 
$m=1/4$ and $m=1/2$.
The energy for $m=1/4$ is exactly known as $4E/N=(E_1-3/4)/2=-1.25595$ (see
table \ref{dpek}),
because the corresponding eigenstate is a simple product state. 
For $m=1/2$ the ground-state energy can be estimated very
accurately to $4E/N=-0.68059(1)$ by finite-size extrapolation.
The table~\ref{dpfs} shows exact diagonalization (ED)
results for $E(m=1/2)$ up to $N=40$ sites.
Obviously, the energy per site increases 
with increasing
chain length for an even number of plaquettes, whereas the energy per site
decreases with increasing chain length for an odd number of plaquettes.
Both energies meet each other at $N \to \infty$.
We note, that 
for the considered maximal odd and even chain lengths of $N=36$ and $N=40$
already the first five digits of $E(m=1/2)$ coincide. 

Second we have to find at least one state with magnetization $m'$ between
$m=1/4$ and $m=1/2$ having an energy below the straight line
connecting the $m=1/4$ and $m=1/2$ points in the $E(m)$ diagram. 
Then with increasing field the system would occupy next after the state with 
$m=1/4$ the state with $m'$ and not that one with  $m=1/2 > m'$. 
According to \cite{ivanov} low-lying states of certain $m$ belong to the
class of product eigenstates, for which in every $(k+1)$-th 
plaquette the spins $\vec{S}_1$ and  $\vec{S}_3$ form 
a vertical dimer singlet $(\vec{S}_1 +\vec{S}_3)^2 = 0$
(i.e., the quantum number $S_{13}$ in Koga's notation is zero). 
The finite strips between two vertical dimer singlets contain
$k$ vertical dimer triplets ($S_{13}=1$).
Since these strips are decoupled from each other we call them 
fragments of length $k$ here where such a fragment
consist of $N_k=4k+2$ spins.
The energies $E$ of those eigenstates can be expressed by the energies of 
the
finite fragments $E_k$ which can be calculated exactly
up to a maximum fragment length of $k=9$ (see table
\ref{dpek}). 
For instance, the energy $E$ 
of states consisting of $N/(4k+4)$ fragments of identical length
$k$ separated by one vertical dimer singlet each 
is $4E/N = (E_k -3/4)/(k+1)$.

From the exact numerical data of table \ref{dpek}
it can be found that all of these 
states have energies below the straight line connecting the 
$m=1/4$ and $m=1/2$ points in the $E(m)$ diagram.
That clearly shows, that we have {\it not} only a single
jump between the $m=1/4$ and the $m=1/2$ plateau.

Let us now present a more detailed analysis of the magnetization process
between $m=1/4$ and $m=1/2$. 
We start from the $m=1/4$ eigenstate, which is a product state of identical
fragments of length $k=1$ each having total spin $S_{k=1}=1$ leading to
a total spin of the chain $S_{chain}= S_{k} \cdot N/(4k+4) = N/8$.   
To increase the magnetization we have to insert one fragment of length
$k=2$  with total spin $S_{k=2}=2$. 
The next eigenstates with higher magnetizations can be generated by replacing
further fragments of length $k=1$ by fragments of length $k=2$.
Due to the product nature of these eigenstates the energy necessary 
to increase the magnetization is constant
and defines the critical field for leaving the $m=1/4$ plateau.
After all fragments of length $k=1$ are replaced by fragments of length
$k=2$ the system has reached a state with 
$S_{chain}= S_{k} \cdot N/(4k+4) = N/6$, i.e., a $m=1/3$ plateau. 
The energy of this $m=1/3$ eigenstate is $4E/N=(E_2-3/4)/3=-1.076056$ 
(see table
\ref{dpek}).
This fragmented structure of the relevant eigenstates 
is also responsible for the finite-size effects, since for finite chains 
the splitting of the
chain in fragments must be compatible with the total length of the chain.
To get higher
magnetizations all fragments of length $k=2$ and $S_{k=2}=2$ 
must now be replaced successively
by fragments of length $k=3$ and $S_{k=3}=3$ and so on.
This scenario generates a series of plateaus at $m=k/(2k+2)$ where
the width of the plateaus decreases rapidly,
because the energy difference between
fragments of length $k$ and $k+1$ is decreasing with increasing $k$.
The critical field between two plateaus can be calculated from the 
energies of the fragments $E_k$
\begin{equation}
 h_c(k)=kE_k-(k+1)E_{k-1}+3/4.
\end{equation}
The energies of the fragments $E_k$ can be found in table \ref{dpek}.
Figure \ref{dpmag} gives an overview about the resulting magnetization
curve. In contrast to the corresponding Fig~.4 in \cite{koga00} 
we get a staircase like infinite series of plateaus.
We emphasize, according to the linear programming theorem (see,
\cite{niggemann} and references therein), fragmented states with nonuniform
fragment lengths have higher energies than corresponding states built 
by fragments of identical length. Furthermore our 
numerical data clearly show 
that the energy of the above considered fragmented states is signi\-ficantly
below the energies of competing low-lying non-fragmented states
(i.e. states where all quantum numbers $S_{13}=1$).

We notice, that our result is in accordance with the general   
rule of Oshikawa {\it
et al.} \cite{oshikawa97} that $n(s-{\bar m})$ has to be an integer.
In our case the period of the ground state $n$ is $4(k+1)$, the spin
$s$ is one half and the magnetization per spin ${\bar m}$ is $k/(4k+4)$
(the magnetization per site $\bar m$ corresponds to on half of the quantity
$m$ used in this comment).

Finally we mention, that the magnetic phase diagram on the (j-h) plane 
presented in Fig.~9 
in \cite{koga00} has to be extended taking into account the additional
plateaus at $m=k/(2k+2)$.

In conclusion, we have found  an infinite series of
magnetization plateaus in a frustrated one-dimensional quantum spin system
which is closely related to the possibility of fragmentation of the
considered chain. Though the possibility of fragmentation has been observed
also for frustrated two-leg spin ladder \cite{mila,honecker}, in the
spin ladder system
only very
simple ground states seem to be relevant for its magnetization curve.
Hence the sequence of plateaus with rational $m(k)=k/(2k+2)$ found in this
paper seems to be a novel property of quantum spin systems, not found so far.


\vspace{0.5cm}
\begin{figure}
\begin{center}
 {\epsfig{file=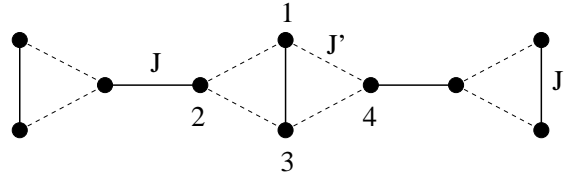}}
\end{center}
\caption{
  The orthogonal dimer chain.
}
\label{fig1}
\end{figure}

\begin{figure}
\begin{center}
 {\epsfig{file=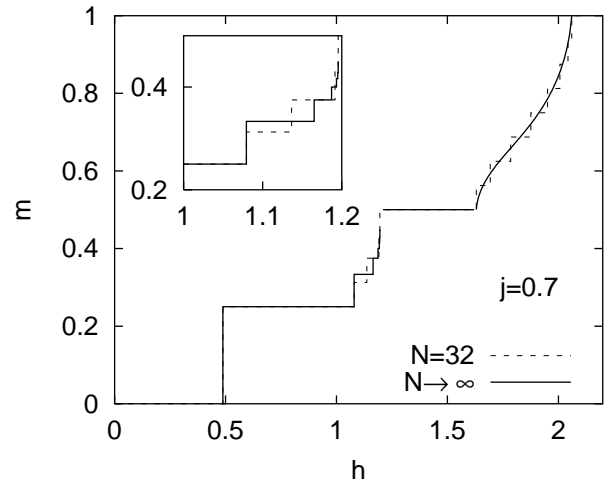}}
\end{center}
\caption{
  Magnetization $m$ of the orthogonal-dimer spin chain for $j=0.7$
  for the thermodynamical limit ($N=\infty$).
  The curve $m>1/2$ is only a guide for the eyes.
  The dashed line is the exact diagonalization result
  of the finite system $N=32$.
}
\label{dpmag}
\end{figure}


\begin{table}
\caption{ Finite-size dependence of the energy 
 of $m=1/2$ states $E(m=1/2)$ for $j=0.7$ up to $N=4N_p=40$ sites.
 $N_p$ is the number of
 plaquettes (vertical dimers) in the chain with periodic boundary
 conditions.}
\label{dpfs}
\begin{center}
\begin{tabular}{cc|cc}
 $N_p$ & $E/N_p$  & $N_p$ & $E/N_p$ \\ \hline
  1 & -0.50000000 &  2 & -0.70313092 \\
  3 & -0.67602844 &  4 & -0.68187420 \\
  5 & -0.68023225 &  6 & -0.68071087 \\
  7 & -0.68055572 &  8 & -0.68060683 \\
  9 & -0.68058916 & 10 & -0.68059535 \\
\end{tabular}
\end{center}
\end{table}

\begin{table}
\caption{ Energies of fragments of the orthogonal-dimer chain of length 
 $k$ and total spin $S_k=k$ for $j=0.7$ and up to $N_k=4k+2=38$ sites.
 }
\label{dpek}
\begin{center}
\begin{tabular}{cc|cc}
 $k$ & $E_k$ & $k$ & $E_k$ \\ \hline
 0 & -0.7500000 & 5 & -4.5289991 \\
 1 & -1.7619002 & 6 & -5.2096756 \\
 2 & -2.4781677 & 7 & -5.8902882 \\
 3 & -3.1658665 & 8 & -6.5708863 \\
 4 & -3.8480439 & 9 & -7.2514810 \\
\end{tabular}
\end{center}
\end{table}


\begin{references}
\bibitem{ivanov} N.B. \ Ivanov, and J.\ Richter, 
      Phys.Lett.A {\bf 232}, 308 (1997);
      J.\ Richter, N.B. \ Ivanov, and J.\ Schulenburg,
      J. Phys.: Condens. Matter {\bf 10}, 3639 (1998)
\bibitem{ueda} K.\ Ueda and S. \ Miyahira, 
      J. Phys.: Condens. Matter {\bf 11}, L175 (1999).
\bibitem{koga00}A.\ Koga, K.\ Okunishi, and N.\ Kawakami,
 Phys.\ Rev.\ B {\bf 62}, 5558 (2000)
\bibitem{niggemann} H.\ Niggemann, G. \ Uimin, and J.\ Zittartz 
      J. Phys.: Condens. Matter {\bf 9}, 9031 (1997)
\bibitem{oshikawa97}
 M.\ Oshikawa, M.\ Yamanaka, I.\ Affleck,
 Phys.\ Rev.\ Letter {\bf 78}, 1984 (1997)
\bibitem{mila}
 F.\ Mila, Eur. Phys. J. B {\bf 6}, 201 (1998)
\bibitem{honecker}
A. \ Honecker,  F.\ Mila and M.\ Troyer, Eur. Phys. J. B {\bf 15}, 227 (2000)
\end{references}
\end{document}